# A Measure for the complexity of Boolean functions related to their implementation in neural networks.

Leonardo Franco
Cognitive Neuroscience Sector - SISSA
2-4 Via Beirut, Trieste, 34014 Italy
(lfranco@sissa.it)

November 9, 2001

## Abstract

We define a measure for the complexity of Boolean functions related to their implementation in neural networks, and in particular close related to the generalization ability that could be obtained through the learning process. The measure is computed through the calculus of the number of neighbor examples that differ in their output value. Pairs of these examples have been previously shown to be part of the minimum size training set needed to obtain perfect generalization in feedforward neural networks.

The main advantage of the proposed measure, in comparison to existing ones, is the way in which the measure is evaluated, as it can be computed from the definition of the function itself, independently of its implementation. The validity of the proposal is analyzed through numerical simulations performed on different feedforward neural networks architectures, and a good agreement is obtained between the predicted complexity and the generalization ability for different classes of functions.

Also an interesting analogy was found between the proposed complexity measure and the energy function of ferromagnetic systems, that could be exploited, by use of the existing results and mathematical tools developed within a statistical mechanics framework, to explore in a more rigorous way the properties of Boolean functions when implemented on neural networks.

We conjecture that the proposed measure could help as an useful tool for carrying a systematic study of the computational capabilities of neural networks, essentially by permitting an easy and reliable classification of Boolean functions. Possible extensions of the work are also discussed.

# 1 Introduction

The complexity of Boolean functions has been studied since a long time [(Shannon, 1949)], being the aim of defining it to have a measure to decide if a problem is easier to solve than another. The previous definition to be operative needs an agreement on the measures of efficiency. These issues have been addressed within the area of circuit complexity [(Wegener, 1987), (Parberry, 1994)], where the complexity of Boolean functions is analyzed in terms of their implementation on different kind of circuits, from those composed with logical AND/OR gates, to feedforward neural networks with threshold or sigmoidal activation functions. Being efficient models for computation and also by their close relationship to neurophysiological models, a lot of work has been done on the study of the computational properties of neural networks [(Parberry, 1994), (Siu, 1995), (Sima, 2001), (Orponen, 1994)], but despite of the effort many aspects of their implementation and general properties remain unclear [(Baum, 1989), (Haykin, 1994), (Lawrence, 1996), (Reed, 1999), (Franco, 2000), (Caruana, 2001)].

Within the area of circuit complexity the complexity of a Boolean function has been defined as the minimum size of the circuit that could implement it, being the size of the circuit measured as the number of nodes that composed it. Also, the depth of the circuit, that is the number of layers or stages in which the computation is performed, is sometimes used as a measure of complexity, as it is more related to the overall computation time [(Wegener, 1987), (Parberry, 1995)]. Many results have been derived for certain classes of functions, as for symmetric and arithmetic ones, where in general, bounds on the size of the circuits to compute the functions are obtained [(Siu, 1995) ,

(Parberry, 1994)] . Moreover, other measures of complexity exists, depending on a different interest, as for example, the total wire length has been recently introduced [(Legenstein, 2000)], as a more adequate measure to sensory processing systems and also related to the construction of VLSI circuits. Another complexity measure was introduced in [(Dunne, 1995)] where the complexity of Boolean functions is defined according to a lazy evaluation approach, and lower bounds are obtained.

The previous mentioned measures have important theoretical and practical interest but, unfortunately, are very hard to calculate as almost all of them are related to the construction of optimal circuits.

In this paper, we propose a measure to estimate the complexity of Boolean functions that could be computed in a very simple and straight way from the definition of the function itself. Our main motivation for this proposal is its possible use to analyze in a systematic way the computational capabilities of neural networks. In next section the proposal is formulated and analyzed, in section 3 numerical studies are performed on neural networks with the aim of validate our proposal. In section 4 a connection is established between the proposed measure and the energy function of ferromagnetic systems together with some related considerations on the structure of very complex functions, to finally present the conclusions and possible extensions of this work.

## 2 A simple measure for the complexity of Boolean functions

A Boolean function, $f$, is a map $f : \{0,1\}^N \to \{0,1\}$, being $N$ the number of bits forming the input. The function $f$ is completely defined when the corresponding output for each of all the $2^N$ possible inputs is determined. We will refer in the rest of the paper by an *example* to one of the possible configurations of the input values together with its output, that is a string of $N+1$ bits.

We propose as a valid measure for the complexity, $\mathcal{C}[f]$, of a Boolean function , $f$:

$$\mathcal{C}[f] \;=\; \sum_{i=1}^{\frac{N}{2}} \mathcal{C}_i[f] \;=\; \mathcal{C}_1[f] + \mathcal{C}_2[f] + \ldots + \mathcal{C}_{\frac{N}{2}}[f], \qquad (1)$$

where $\mathcal{C}_i[f]$ is the complexity term of order $i$, taking into account pairs of examples at a hamming distance $i$. For example the first term, $\mathcal{C}_1[f]$ takes into consideration pairs of examples differing in just one input bit (nearest neighbor examples), the second term examples differing in two input bits, and so on. Along the paper we mostly use the first term, with $i = 1$, being equal to:

$$\mathcal{C}_1[f] = \frac{1}{N_{ex} * N_{neigh}} \sum_{j=ex} \left( \sum_{nn.\ ex.} (|T(ex_j) - T(nn.ex.)|) \right), \qquad (2)$$

where the first factor, $\frac{1}{N_{ex} * N_{neigh}}$ is a normalization one, counting for the total number of pairs considered, $N_{ex}$ is the total number of examples equals to $2^N$, and $N_{neigh}$ stands for the number of neighbor examples at a hamming distance of 1, as we are considering the first term of the series. The indicated sums are performed first, over all the examples, and for every one of the examples, over all the $N$ existing neighboring examples, denoted in Eq. 2 by $nn.\ ex.$. $T(ex_j)$ indicates the target (output) of the example being considered, while $T(nn.ex.)$ refers to the output of a neighboring example. The rest of the terms, $\mathcal{C}_i[f]$, with $i = 2, \ldots, int\left(\frac{N}{2}\right)$, have all the same structure of the first term, i.e., consist of a normalization factor, and two sums, over all the examples and for each one of them over the neighboring ones at a hamming distance of $i$. Note that each term is normalized independently, being the normalization constant equals to the total number of pairs that exists at the hamming distance considered. Clearly, this normalization permits a maximum value per term of 1, but in practice and due to the structure of the examples it is not possible that the terms take this maximum value, except for the first term. For example, there no exists a function for which all the second nearest neighbors (hamming distance = 2, between the inputs) have opposite outputs, as if for some examples their second nearest neighbors have opposite output, the same condition implies that exists other examples for which this condition does not hold. In this way, the definition of complexity essentially consists on counting the number of pairs of neighboring examples having different outputs and weighting them according to the distance between the examples forming the pair.

The fact that the definition has $int\left(\frac{N}{2}\right)$ terms and not $N$ terms as it could be, has the justification that till the $int\left(\frac{N}{2}\right)$ term there is apparently no need of an extra correction factor, as the effect of one example is decreasing in value depending on the distance till the $int\left(\frac{N}{2}\right)$ term. Along this work we do not consider higher order terms than the second as it was not necessary and also because we want to keep the definition and calculus of the measure as simple as possible. When it is not precisely specified, the value of the complexity of functions refers to the first term only, i.e. $\mathcal{C}[f] = \mathcal{C}_1[f]$.

The presented measure of complexity is inspired on the results obtained in [(Franco, 2000), (Franco, 2001)], where a close relationship between the number of examples needed to obtain valid generalization and the number of neighboring examples with different outputs was found. In those works a method for selecting examples in order to improve the generalization ability has been developed, and through its application to different functions and architectures scaling properties for the generalization ability of feedforward neural networks have been obtained.

The idea behind the proposed measure is presented in a graphical form in Fig. 1, where the AND and XOR functions are shown for the case $N = 2$. The correspondent output values are indicated by empty (0) and filled (1) circles besides the inputs, that are also indicated by a pair of numbers. Different regions containing different outputs are separated by lines (hyperplanes in higher dimensions), as it is usually done to illustrate the idea of linear separable functions [(Hertz, 1991), (Reed, 1999)]. In the figure, lines with arrows in their ends, and crossing the separating hyperplanes are drawn to indicate those neighboring examples that have a different output. The value of the proposed measured of complexity is proportional to twice the number of these pairs of examples with different outputs, through a normalization factor equals to the total number of examples times the number of neighbors that each example has. In the presented cases of Fig. 1, the total number of examples is 4, with every example having two neighboring ones, obtaining a complexity equals to 0.5 for the AND function, and 1, for the XOR function.

The proposed measure, in first approximation (only first term considered), is simple to calculate, being a sum of $(N * 2^N)$ binary terms, permitting to classify $2^{2^N}$ existing functions of $N$ arguments in $(N * 2^N)$ different complexity categories. An interesting point is that its calculus does not depend on the construction of optimal circuits, being independent of any device or platform used to implement the functions.

## 3 Validation of the proposal

In order to validate the usefulness of the proposed measure, different simulations were performed on feedforward neural networks, to analyze how important features like generalization ability and training time are related to the measure. The simulations were done on one hidden layer architectures, fully connected between adjacent layers, applying backpropagation as learning algorithm. The choice is based on the fact that this combination of architecture plus algorithm, is the most widely used, being successfully applied to a large variety of problems [(Haykin, 1994), (Reed, 1999)].

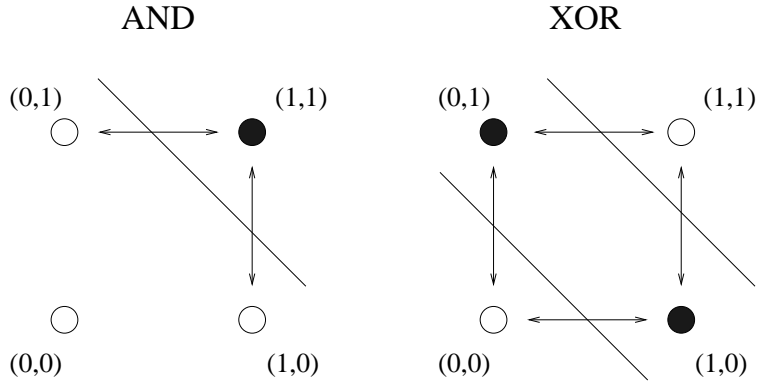

Figure 1: Schematic drawing of the idea behind the complexity measure proposed in this work, exemplified on the AND and XOR functions for the case $N = 2$. The color of the circles, (white=0, black=1), indicates the output value of the inputs, that are also indicated by a pair of numbers. The complexity of the functions is proportional to the number of pairs of neighboring examples having opposite outputs, and these pairs are indicated in the figure by segments ending with arrows.

We focus our attention on the calculus of the generalization ability and on the time needed to train the system, without analyzing the computability of the implemented functions by the used architecture. This last mentioned task is a very complicated one, demonstrated to belong to the class of NP-complete problems [(Judd, 1990)], but very interesting issue, that deserves its own study. Even more, we hope that the present work could serve as a useful tool to carry on a systematically study of the computational capabilities of neural networks.

For the implementation of the training process, the whole set of examples is split in 3 groups: training, validation and test sets. The size of these sets was kept constant in proportion to the number of total examples existing for each architectures: the training set contains half of the total number of examples selected at random, and the validation and generalization test sets, have a fourth of the remaining examples each. It is considered, in a general sense, that the validation set form part of the training set, as in some cases (not in this work) the network could be retrained including these examples. The training procedure consists of training the network using the training set by backpropagation, while monitoring the value of the error on the validation set, to find the point where its lower value is achieved. At that point the generalization error is measured using the generalization set of patterns. It was shown that through this procedure a better generalization could be achieved [(Haykin, 1994), (Reed, 1999)]. Also, we calculate the computation time, measured in epochs, needed to reduce the training error to 0.25. In general, except where indicated, the obtained values are expressed as average results for group of functions with a complexity around the indicated value. Generalization ability and training error figures are given as a fraction of the correct or wrong classified examples, respectively. To be more precise, the generalization ability is always and only calculated on examples that have been never seeing by the network.

We run simulations for all Boolean functions for the case of networks with $N = 4$ inputs, and for symmetric and random generated functions for the case with $N = 8$ inputs, using a variable number of neurons in the hidden layer. In the next subsections, characteristics and results of these simulations are described.

## 3.1 Boolean functions with $N = 4$ inputs: an exhaustive study.

In this subsection, we analyze all 65536 Boolean functions that exists for the case $N = 4$, implementing them on a one hidden layer network with 7 units. The question, whether the used structure could compute or not a determined function, is not analyzed here but from simulations performed on different network sizes, could be infer that a high proportion of them can be computed by the used architecture. The lower values obtained for the training error at the end of the training process (750 epochs), shown in the last column of table 1, reinforce the previous hypothesis. We also calculate the generalization ability and the computational time needed to reduce the training error down to 0.25. The results are presented in table 1, where we group them according to the complexity of the functions within the ranges indicated in column 1.

Table 1: Some properties obtained through a learning process for *all* the Boolean functions with $N = 4$ inputs, implemented in a feed-forward 4-7-1 neural network architecture. The complexity measure is calculated using only the first order term, i.e. $\mathcal{C}[f] = \mathcal{C}_1[f]$ .

| Complexity range ($1^{st}$ order) | Number of functions | Generalization ability | Train. time (epochs) | Final train. error |
|---|---|---|---|---|
| $0.0 - 0.1$ | 2 | 1.00 | 1 | 0.000 |
| $0.1 - 0.2$ | 96 | 0.88 | 13 | 0.005 |
| $0.2 - 0.3$ | 424 | 0.83 | 42 | 0.005 |
| $0.3 - 0.4$ | 8416 | 0.65 | 67 | 0.005 |
| $0.4 - 0.5$ | 13568 | 0.55 | 88 | 0.005 |
| $0.5 - 0.6$ | 34092 | 0.46 | 114 | 0.011 |
| $0.6 - 0.7$ | 8416 | 0.41 | 137 | 0.017 |
| $0.7 - 0.8$ | 424 | 0.40 | 158 | 0.022 |
| $0.8 - 0.9$ | 96 | 0.35 | 195 | 0.029 |
| $0.9 - 1.0$ | 2 | 0.50 | 208 | 0.062 |
| Total or average | 65536 | 0.50 | 102 | 0.009 |

From the results, displayed in table 1, it is possible to appreciate a high correlation between the values of the proposed measure of complexity and the practical features calculated to test the measure. The computed quantities behave as expected, given that for an increasing complexity of the functions the generalization ability decrease monotonically, and the computational training time and final training error are larger for more complex functions. Only for the two more complex functions [1] in the range $0.9 - 1$, the generalization ability value is higher than those obtained for less complex functions, fact that it is not very significant as the value is calculated only on two functions. We also calculate the values of the features shown in table 1, for the case of including the second order term in the complexity measure, i.e. $\mathcal{C}[f] = \mathcal{C}_1[f] + \mathcal{C}_2[f]$, obtaining similar results, in general terms, and that is why we do not show them. For the case of considering the second order term, the generalization ability and training times show even a better agreement with the complexity measure, as expected for a more accurate measure, evidenced by a less dispersion around the mean values, that we measured for both cases through the calculus of the standard deviation. The obtained values for the standard deviation, averaged across all functions, were 0.238 and 0.230 for $\mathcal{C} = \mathcal{C}_1$ and $\mathcal{C} = \mathcal{C}_1 + \mathcal{C}_2$ respectively.

It is worth noting that the total average generalization ability obtained is 0.50, property that also holds in general cases whenever the average is done on the whole set of Boolean functions. In subsection 3.4 we demonstrate this property and discuss some implications arising from it.

## 3.2 Symmetric functions

An important class of Boolean functions are the symmetric ones, those with values independent of the order of the input, i.e., the output depends only on the total number of input bits ON (the number of 1's). The class of symmetric functions contains many fundamental functions like sorting and all types of counting ones, including also the well known parity function [(Wegener, 1987), (Parberry, 1994), (Siu, 1995), (Dunne, 1995), (Reed,1999)]. They have been extensively studied and many lower bounds have been obtained for circuits computing them. A general result states that a circuit of size $\mathcal{O}(\sqrt{N})$ gates and depth 3 with polynomial weights is enough to compute all the symmetric functions [(Siu, 1991)]. The number of symmetric functions, for $N$ input bits, is $2^{N+1}$, since the set of input examples can be divided in $N + 1$ groups, with 0 to $N$ activated inputs.

---

[1] these two functions are the parity and its opposite function (NOT parity)

We carry on simulations with all the 512 symmetric functions for the case $N = 8$, using neural networks with one hidden layer with a variable number of hidden neurons within the range from 2 to 29. The results for the generalization error and computational time, are plotted vs. the complexity (first order) in Figs. 2a and 2b. Certain variability exists for functions with the same complexity, but on an average sense, a good agreement is found, as for an increasing level of complexity the generalization ability diminishes, while the training times increase. In Fig. 3 the generalization ability for symmetric functions is plotted vs. the number of hidden neurons in the architectures used, for all the functions grouped in intervals of 0.1 of the complexity (first order). The correlation of the generalization ability to the complexity is clearly preserved for different network sizes, as no significant variation is observed with the number of hidden units, except for some changes in the generalization ability when the network size increases from smaller values. This effect seems to be related to the computational capabilities of the architectures, that are expected to vary much when the size of the network is increased starting from small networks.

It has been observed that a better generalization ability could be obtained, for certain functions, when implementing them on very large network architectures (see [(Lawrence, 1995), (Caruana, 2001), (Franco, 2002)]). Respect to this effect, and related to Boolean functions, it seems that this occurs only for very low complexity functions, those with a complexity lower than 0.3 (see Fig. 3), as in all other cases optimal network size values were found. An interpretation of this result is that increasing the size of the network increase much more the number of configurations that implement low complexity functions than those implementing more complex ones.

It is also shown in Fig. 4a and Fig. 5 the behavior of the generalization ability vs. the complexity for symmetric functions when using the first and second order complexity measure, in comparison to the results obtained for random generated functions (these results are discussed in next subsection when the random generated functions are analyzed). Also in Fig. 4b the values for the training time versus the complexity (first order) are plotted for the mentioned classes of functions.

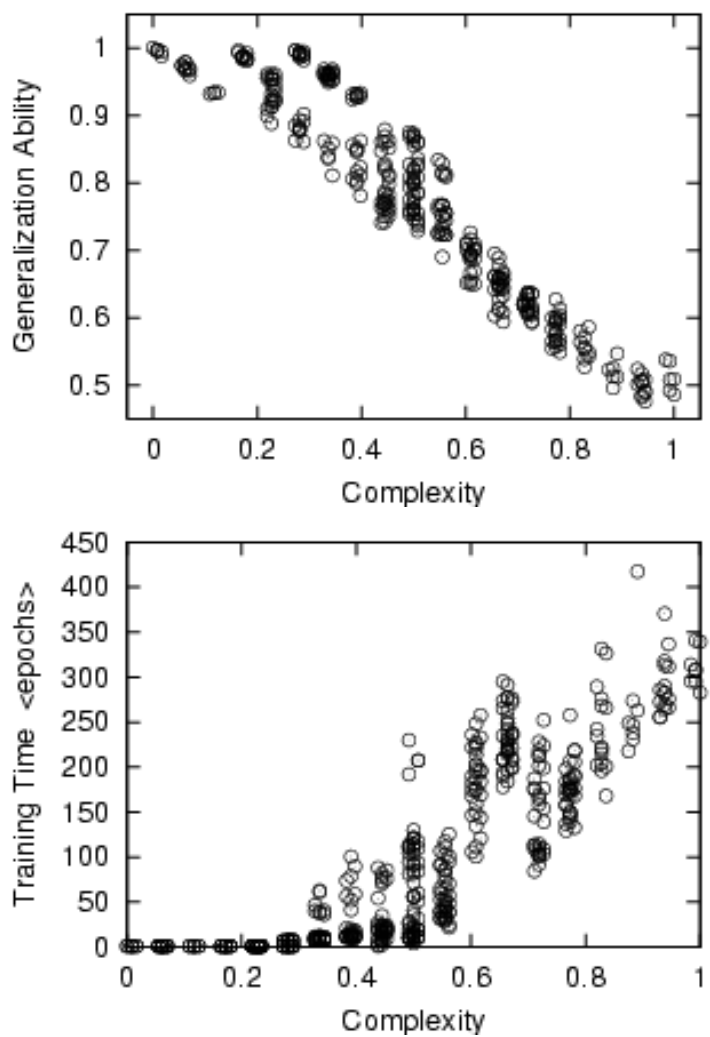

Figure 2: Generalization Ability (a), and Training Time (b) vs. Complexity ($1^{st}$ order) for all the symmetric functions with $N = 8$, in a 8-29-1 neural network architecture. The generalization error is measured through a validation procedure, in which the net is trained on half the total number of examples, 128, while monitoring the value of the error on other 64 examples, and at the best validation point the generalization error is measured on the remaining 64 examples. The training time is measured at the point when the training error is reduced to 0.25.

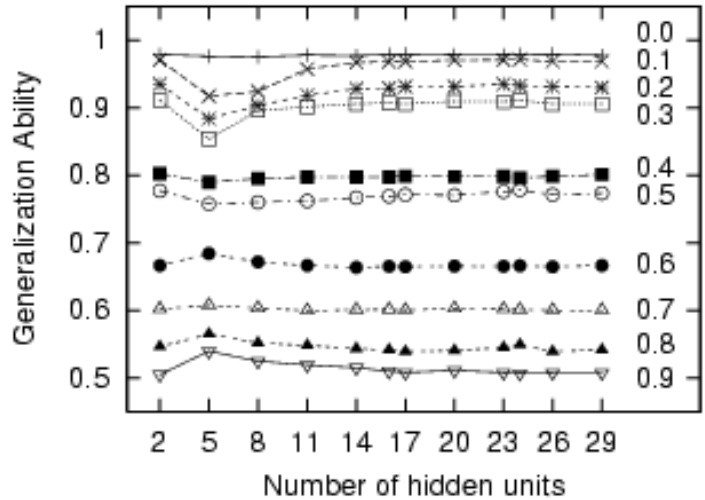

Figure 3: Generalization ability vs. number of neurons in the single hidden layer of the neural network architectures used to compute all the symmetric functions with $N = 8$ inputs. The different curves plotted in the figure correspond to average values of functions grouped according to their complexity ($1^{st}$ order term considered) in intervals of 0.1. Complexity values for the groups are indicated on the right side of the figure.

## 3.3 Random functions

There exists $2^{2^N}$ Boolean functions of $N$ inputs, making their exhaustive study very complicated except for very simple and small cases, like the case $N = 4$ studied in a previous subsection. The problem of generating sample functions to represent the total distribution of functions is not very simple, as for example, trying to generate random functions by a random assignment of outputs ends with a set of functions, all with a complexity (order one) around 0.5, as the probability of obtaining a function with a different complexity is very low. Despite the lower probability that functions different to random generated ones have, considering the whole set of Boolean functions, it is important to analyze different functions within different ranges of complexity, as many practical problems seems to be not very complex (like many pattern recognition problems), and even in certain cases for complex ones a high degree of internal structure exists between their input values (for example, the case of the parity function) making these functions quite different in comparison to random ones.

Considering the first order complexity measure, up to first nearest neighbors, the most difficult function is parity (and also its opposite counterpart), that has the property that for any example all the neighbor ones have a different output. A lot of work has been devoted to the study of the parity function [(Thornton, 1996), (Franco, 2001)], both for historical reasons and because it is considered a very difficult function. For our study, we use the parity function as an initial configuration to generate very complex functions. For the case $N = 8$, we generate two sets of random functions: Random 1 and Random 2 sets. The set Random 1 is generated in two parts: the simpler functions are generated by modification of the constant functions, those having all output values, equal all to 0 or to 1. Thus, we take the initial configuration and modify the output values randomly with a certain probability, to obtain functions with a complexity (order 1 considered) from zero to 0.5. The other part, covering the range of complexities from 0.5 to 1 is generated in the same way but taking as initial configuration the parity function.

The Random 2 set, is generated straightforward in a single way process, starting from the parity function as initial configuration, and by flipping with certain probability only the bits that are equal to 1, in the definition of the parity function. In this way, we obtain functions similar to the parity function when the probability of flipping bits is very low, but when this probability of flipping increase to higher values functions similar to the constant function are obtained, and thus, functions that cover the whole range of complexities (order 1) from 1 to 0 are generated.

It seems intuitively that the set Random 1 is more general and more representative of the distribution of functions, than the set Random 2, as the functions belonging to the first set have less internal structure.

In Figs. 4a and 4b we show the results obtained for the two sets of random generated functions and also for comparison, those corresponding to symmetric functions are also shown. Within each class of functions a good agreement between the complexity measure and the calculated feature (generalization ability and computation time) is obtained, but between different classes we observe that for functions with the same complexity a different generalization ability in average is obtained.

Trying to obtain a better matching between the generalization ability for the different classes of functions, we calculate the complexity measure but including the second order term of the expansion (see Eq. 1), i.e. $\mathcal{C} = \mathcal{C}_1 + \mathcal{C}_2$. The calculus now involves up to pairs of second order nearest neighbor and some more computational cost, as there exist $N * (N-1)/2$ second nearest neighbors for each example. In Fig. 5 the results are shown for the two sets of random generated functions Random 1 and 2 and also for the symmetric functions. As it could be seen from the figure, a good agreement is obtained for the generalization ability values for functions with the same complexity for both sets of Random functions. In comparison to the results obtained for symmetric functions some small difference remains, but in our opinion, this reflects the fact that the number of symmetric functions is quite small, making possible that some special characteristics of the small set appears, as for example, note the high value for the generalization ability for functions with a complexity around 0.6.

## 3.4 A general result about average generalization and the existence of very complex functions

From a comparison of the results obtained for all the functions with $N = 4$ and those for random generated functions with $N = 8$, it could be noticed a clear difference in the values of the average generalization error. While for the $N = 4$ case, this value was found to be 0.5 (see Table 1), for the case $N = 8$, a value greater than 0.5 would be obtained, as a consequence that for all the considered functions the generalization error is greater than 0.5 (see Figs. 4a and 5.). We do not calculate this value because the calculus would involve using the probability distribution of functions, according to their complexity, that we do not know.

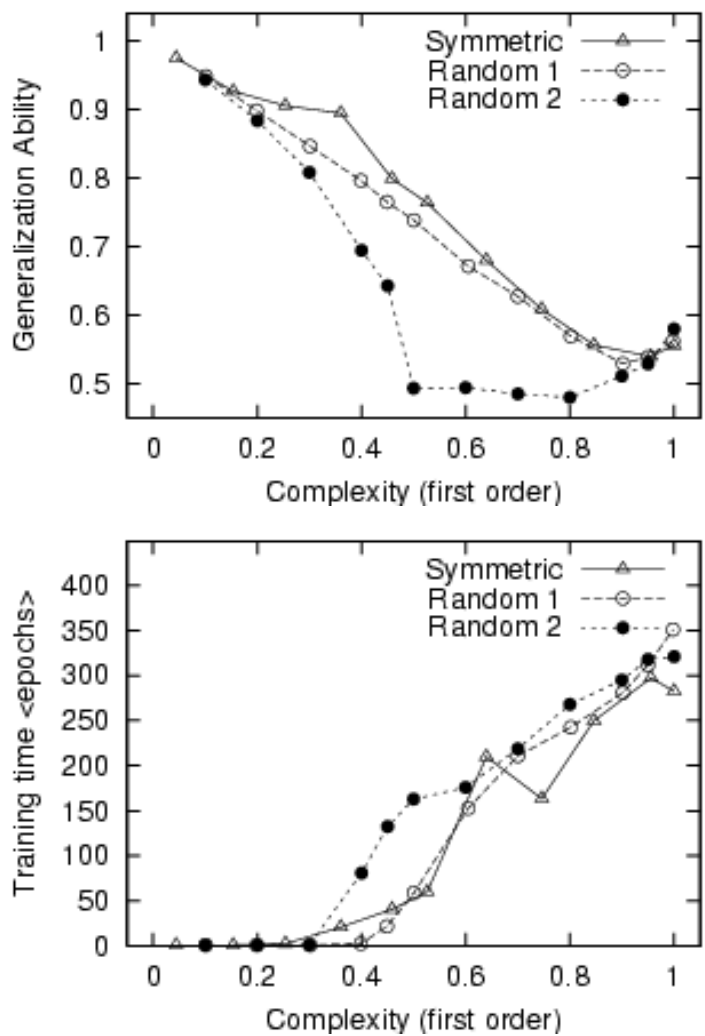

Figure 4: Generalization Ability (a) and training time (in epochs) (b) vs. Complexity ($1^{st}$ order) for three different generated families of Boolean functions with $N = 8$ inputs (see text for details).

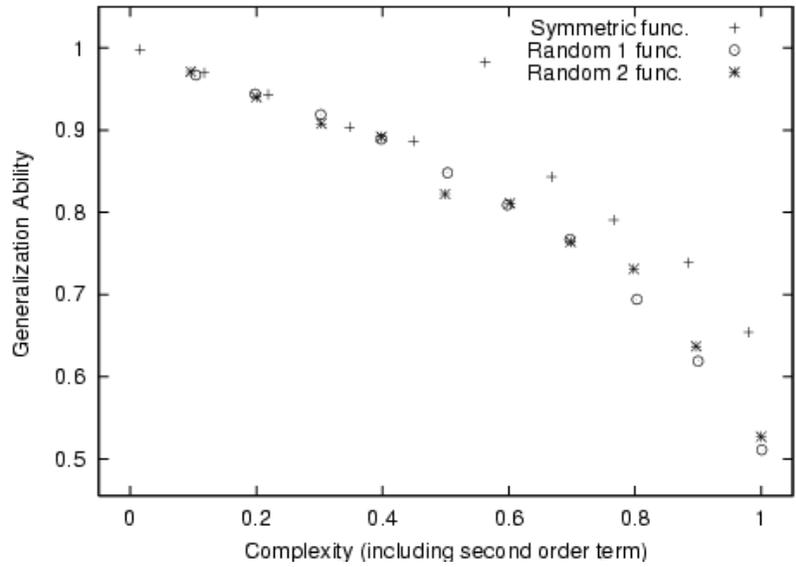

Figure 5: Generalization Ability vs. Complexity ($1^{st} + 2^{nd}$ order terms considered) for three different generated families of Boolean functions: Random1, Random 2, and Symmetric functions (see text for details).

To solve this apparent controversy, we realize that the fact of having obtained a value 0.5 for the generalization ability of all Boolean functions for the case $N = 4$ is a general property when the average is computed on the whole set of Boolean functions is taken into account, result that we state as:

*For any fixed architecture device, that could be trained by examples, as for example a neural network, the average generalization error on a* complete *set of Boolean functions is* 0.5 *independent of the learning algorithm used.*

To demonstrate the previous assertion, we consider the case in which the network is trained on $2^N - 1$ examples, and the generalization ability is tested on the remaining example, but the procedure could be easily generalized to other cases. All the possible training sets, $2^N$, are considered assuming that the network is trained only once on each different training set, but again this is not a restriction for the demonstration.

Thus, the average generalization error could be written as:

$$\mathcal{E}_g = \frac{1}{2^{2^N} 2^N} \sum_{\text{all functions}} \sum_{\text{all training sets}} |Target_i - Output_i|, \quad (3)$$

that we rearrange as a sum over pairs of functions sharing $2^N - 1$ training examples (input and output), but differing in the remaining one, and now the generalization error can be written as:

$$\mathcal{E}_g = \frac{1}{2^{2^N} 2^N} \sum_{\text{coupled pairs}} |Output_i - 1| + |Output_{\bar{i}} - 0| \quad (4)$$

where $i$ indicates an example and $\bar{i}$ denotes another example sharing the same training set. The value of $|Output_i - 1| + |Output_{\bar{i}} - 0|$ is 1 (for all the terms) because the training set is the same for both cases but, the test examples for the pair, in which the generalization ability is tested, have opposite values. Eq. 4 contains half the number of terms of Eq. 3, as we grouped the terms in pairs, having exactly $\frac{2^{2^N} 2^N}{2}$ terms equal to 1, half of the normalization factor, and thus the value 0.5 for the average generalization error is obtained.

A consequence arising from the previous result, is that some functions are not being considered for the case $N = 8$, and also that these functions are very complex ones with an expected average generalization ability lower than 0.5.

From an analysis of the way in which the functions used in the case $N = 8$ were generated, it is possible to see which are the functions that have not been considered. This "new" set could be generated, again by starting from the parity function, but by flipping not individual output bits, but strings of contiguous output bits when the examples are ordered according a certain way, for example according the number of inputs bits ON, i.e. regions or domains are created where the function seems to be the parity function, but in another region the outputs are equal to those of the NOT parity function. Through the mentioned process we obtain functions with a measure of complexity (including the second order term) greater than 1, in general within the range from 1 to 1.3 (for the case $N = 8$). In next section we analyze through the introduction of an statistical mechanics analogy, the structure of these very complex functions and also we show in Fig. 6 an example of the structure of one of the highest complex functions for the cases $N = 3$ and $N = 4$.

To analyze how the generalization ability of this set of very complex functions depends on the size of the network, we run simulations on different architectures, varying the size of the hidden layer, obtaining in all cases an average generalization ability below 0.5. In table 2 the generalization ability, the training time in epochs and the training error at the end of the training process are shown for feedforward neural networks with $N = 8$ inputs an variable number of hidden neurons, from 1 to 45, for functions with an average complexity (considering the second order term) around 1.1.

# 4 An analogy between the proposed complexity measure and the Hamiltonian of a spin system.

A close look to Eq. 2, where the first order term of the complexity measure is defined, permits to see an exact correspondence between this equation and the Hamiltonian of a system of spins with ferromagnetic interactions between first nearest neighbors spins [(Cannas, 1996)]. The other terms in the expansion of Eq. 1, those for $i = 2, \ldots, \frac{N}{2}$, represent the ferromagnetic interaction between second, third, etc., nearest neighbors spins in a lattice with a particular architecture defined by the structure of the inputs of the Boolean functions, and with an interaction constant depending on the distance between spins and whose value is related to the normalization factor presents in all the terms of the expansion (ref. Eqs. 1 and 2 and section 2). The structure in which the system is defined, could be characterized by the number of neighbors sites as a function of the hamming distance between sites and it is known as the hypercubic cell. It has been introduced by Parisi et al. [(Parisi, 1991)] to analyze the behavior of a spin glass system, being the generalization to N-dimensions of the unit cubic cell, that for the case $N = 2$ corresponds to a square, while for $N = 3$, corresponds to a cube.

Table 2: Generalization ability, training time (epochs) and final training error for very complex functions with an average complexity (second order) around 1.1 implemented on neural networks architectures with $N = 8$ inputs and one hidden layer with a number of hidden neurons between 1 and 45

| Hidden Neurons | Generalization ability | Train. time (epochs) | Final train. error |
|---|---|---|---|
| 1 | 0.469 | 16 | 0.536 |
| 5 | 0.459 | 331 | 0.192 |
| 10 | 0.487 | 437 | 0.003 |
| 15 | 0.452 | 316 | 0.002 |
| 20 | 0.474 | 511 | 0.002 |
| 25 | 0.477 | 314 | 0.002 |
| 30 | 0.469 | 437 | 0.002 |
| 35 | 0.460 | 268 | 0.004 |
| 40 | 0.467 | 533 | 0.002 |
| 45 | 0.463 | 383 | 0.003 |

By using of the presented analogy, it is possible to see that the most complex functions according to the complexity measure defined in Eq. 1, correspond to configurations with maximal energy of this ferromagnetic system, while the simplest Boolean functions (the two constant ones with all the outputs equal to 0 or to 1) correspond to states with the lowest energy defined on the hypercubic cell. Moreover, completely random generated functions with a complexity (order 1, i.e., $\mathcal{C} = \mathcal{C}_1$) of 0.5 (see subsection 3.3), correspond to states with maximum probability for a ferromagnetic system at very high temperatures.

For the case $N = 2$, the complete definition of the complexity measure involves only the first term (ref. Eqs. 1 and 2), and the XOR function (see Fig. 1) is the highest complexity function. When the dimension of the system is larger than 2 ($N > 2$), other high order terms are included in the definition of complexity, for example, two terms for the cases $N = 3$ and $N = 4$, and, the generalization to higher dimensions of the XOR function, the well known parity function [(Thornton, 1996),(Franco, 2001)], it is not more the highest complex function, despite of still being a very complex one. In Fig. 6 we show for the cases $N = 3$ and $N = 4$ one of the most complex functions for each case. In the figure, the input values are represented by the coordinates values of the corners of the cubes, also indicated by $(I_1, I_2, \ldots, I_i :$ output), where $I_i$ are the input values, while the color of the spheres, drawn on each corner, reflects the output value for each example. For the case $N = 3$ (left part of Fig. 6), one of the highest complexity functions is displayed (the others functions with equal complexity are obtained through a rotation of the presented one) can be seen as consisting of two XOR functions defined on 2-dimensional squares, corresponding to the right and left faces of the cube, but noting that neighboring examples belonging to different faces (left or right) have the same output. In the same figure, for the case $N = 4$ (right part of Fig. 6), one of the highest complexity functions is represented on an hypercubic cell for the case $N = 4$, that we illustrate as two cubes linked by connections between corners separated by a hamming distance of 1. It is easier to see that within each cube a parity function is defined, but as in the $N = 3$ case, neighboring inputs, belonging to different cubes, have the same output. In a physical language, we refer to these spin configurations, defined on each cube, as anti-ferromagnetic domains.

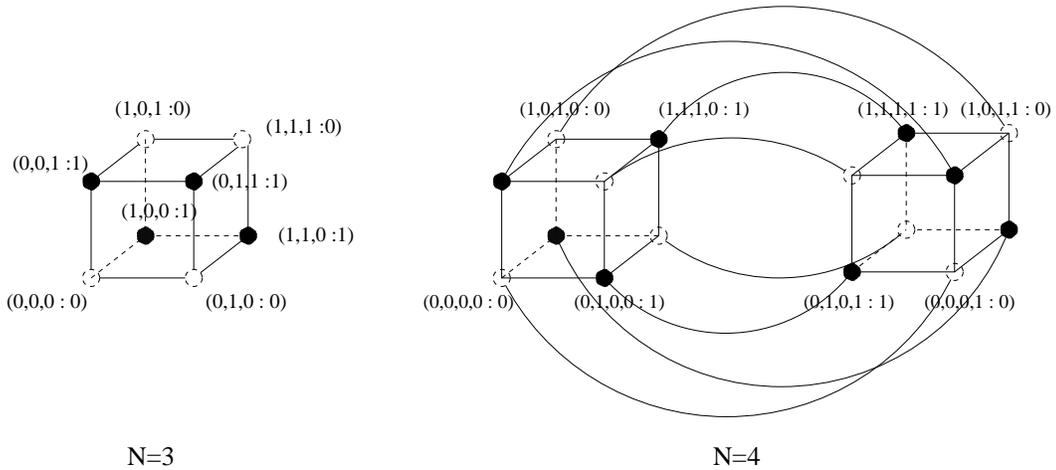

Figure 6: A way to visualize the highest complex functions according to the proposed complexity measure ($2^{nd}$ order considered), for the cases $N = 3$ and $N = 4$. Input examples of the functions are indicated in the way ($I_1, I_2, ..., I_i$ : Output) where $I_i$ are the input values. For all the examples the output values are indicated by the color of the spheres drawn on each corner. In the case $N = 4$ connections between the two cubes link first nearest neighbors, in a way to represent the fourth dimension, in which every example has 4 first nearest neighbors (see also text for more details).

The established analogy could be useful, as many tools and results from statistical mechanics, (see for example [(Cannas, 1996), (Parisi, 1991)], can be applied to obtain more rigorous and complementary results. We are currently analyzing this connection within a more mathematical framework, and the results will be published elsewhere.

# 5 Conclusions

From ideas taken from the calculus of the minimum number of examples needed for perfect generalization, applied to different functions and architectures [(Franco, 2000), (Franco, 2001)], a simple measure permitting to classify Boolean functions according to their complexity was proposed (Eqs. 1 and 2). The main advantage, compared to other existing measures, like those existing within the area of circuit complexity [(Parberry, 1994)], is the fact that the introduced measure could be easily computed from the definition of the function itself, independently of its implementation.

The measure was tested on the whole set of Boolean functions for $N = 4$, and on different classes of Boolean functions, like symmetric and random generated ones, for the case $N = 8$, for different network sizes. In all cases a strong correlation was obtained between the proposed measure and practical features at the time of the implementation of the functions on neural networks, as generalization ability and training times. On average, the generalization ability decrease monotonically with the complexity of the functions, while the training times increase, as it is expected for more complex functions. Moreover, for the case when all the symmetric functions with $N = 8$ inputs were considered, the generalization ability and training time show a nice agreement with the proposed measure, for almost *all* functions, as for functions with the same complexity similar values of the generalization ability and training times were obtained (ref. Fig. 2).

The proposed measure consists in different terms accounting for the effect of pairs of neighbor examples with different outputs, weighted by the distance between the examples. For functions belonging to the same class (according the way in which the functions are generated, see section 3 for details), the first order complexity measure, $\mathcal{C}_1$, was enough to establish a nice correspondence between the generalization ability and the complexity value proposed, but when the comparison was done for functions with the same complexity value but belonging to different classes (see Fig. 4), some discrepancies appeared, making necessary to use the second order term of the series, $\mathcal{C}_2$, to obtain similar generalization values between different groups of functions (see Fig. 5). The previous results indicate that sometimes, depending on the scope of the analysis, higher order terms have to be considered, making the calculus of the complexity more computational demanding but without affecting the computability of the measure as its calculus involves very simple operations.

When the whole set of Boolean functions, for the case of $N = 4$, was analyzed, an average value of 0.5 was obtained for the generalization ability, result that then it was extended to any input size. We proved that this property is a general one, valid whenever the average of the generalization ability is calculated on the whole set of Boolean functions. Even more, the last mentioned result led us to discover the existence of a set of high complex functions, for which a generalization ability below 0.5 was predicted and that we tested on different architectures.

Also an analogy was established between the complexity measure and the Hamiltonian of a system of spins with ferromagnetic interactions, fact that can help to the understanding of the complexity structure of the Boolean functions, as results and tools developed within an statistical mechanics framework can be used.

A question, related to the proposed measure, that arise is whether the presented complexity measure has any relationship with other complexity measures related to Kolmogorov complexity [(Cover, 1991), (Lovász, 1996)]: in a general sense, the proposed complexity has a relationship, as the most complex functions are the less predictable ones, those for which a low generalization ability is obtained. Considering a function as a string of the $2^N$ output bits, as it would be the case of applying measures related to the Kolmogorov one, would be less accurate than the proposed measure, as the measure introduced in this work takes into account the structure of the space of examples, by establishing a measure of the influence of pairs of examples on the generalization ability, related to the hamming distance between the examples, and thus having, in a general sense, more information about the complexity of the functions.

In light of the results, we conjecture that the complexity measure proposed in this work could be a valuable tool for exploring the complex computational capabilities of neural networks. We are currently exploring the extension of our proposal to continuous functions, more frequently found in real problems, together with a more detailed analysis of the properties of the proposed measure within a more mathematical framework, trying to exploit the analogy between the proposed measure and the energy function of magnetic systems.

# Acknowledgments

L.F. thanks Roberto Zecchina for fruitful discussions and Sergio A. Cannas for a critical reading of the manuscript. Partial support from Human Frontiers Program Grant RG0110/1998-B is acknowledge.

# References

Baum, B.E. What size of neural net gives valid generalization ? *Neural Computation*, *1*, pp. 151-160. (1989).

Cannas, S.A. and Tamarit, F.A. Long-range interactions and nonextensivity in ferromagnetic spin models. *Phys. Rev. B*, *54*, pp. R12661-R12664. (1996).

Caruana, R., Lawrence, S. and Lee Giles, C. Overfitting in Neural Networks: Backpropagation, Conjugate Gradient, and Early Stopping. *Advances in Neural Information Processing Systems*, Denver, Colorado, MIT press, *13*, pp. 402-408. (2001).

Cover, T.M. and Thomas, J.A. *Elements of Information Theory*. Wiley, New York. (1991).

Dunne, P.E., Leng P.H., and Nwana, G.F. On the Complexity of Boolean Functions Computed by Lazy Oracles. *IEEE Transactions on Computers*, *44*, pp. 495-502. (1995)

Franco, L. and Cannas, S.A. Generalization and Selection of Examples in Feed-forward Neural Networks. *Neural Computation*, *12*, pp. 2405-2426. (2000).

Franco, L. and Cannas, S.A. Generalization properties of modular networks implementing the parity function. *IEEE Transactions in Neural Networks*, *12*. (2001) (In press).

Franco, L. and Jerez, J.M. Modularity: a natural choice for improving generalization. To appear in: *Proceedings of WSES'2002 International Conference on Neural Networks Applications*. (2002).

Haykin, S. *Neural Networks: A Comprehensive Foundation*. Macmillan. (1994).

Hertz, J., Krogh, A., and Palmer R. G. *Introduction to the Theory of Neural Computation*. Addison-Wesley, Redwood City, CA. (1991).

Judd, J.S. *Neural Network design and the complexity of learning.* MIT Press. (1990).

Lawrence, S., Giles, C. L., and Tsoi, A. C. What size neural network gives optimal generalization ? Convergence properties of backpropagation. In: *Technical Report UMIACSTR -96-22 and CS-TR-3617*, Institute for Advanced Computer Studies, University of Maryland. (1996).

Legenstein, R.A., and Maass, W. Foundations for a circuit complexity theory of sensory processing. *Advances in Neural Information Processing Systems*, Denver, Colorado, MIT press, *13*, pp. 94-110. (2001).

Lovász. L. Information and complexity (how to measure them?) In: *The Emergence of Complexity in Mathematics, Physics, Chemistry and Biology* (ed. B. Pullman), Pontifical Academy of Sciences, Vatican City, Princeton University Press, pp. 65-80. (1996).

Orponen, P. Computational complexity of neural networks: A survey. *Nordic Journal of Computing*, *1*, pp. 94-110. (1994).

Parberry, I. *Circuit Complexity and Neural Networks.* MIT Press. (1994).

Parisi, G., Ritort, F., and Rubí, J.M. Numerical results on a hypercubic cell spin glass model. *J. Phys. A Math. Gen.*, *24*, pp. 5307-5320. (1991).

Reed, R.D., and Marks, R.J.II. *Neural Smithing: Supervised Learning in Feedforward Artificial Neural Networks.* Cambridge, MA. (1999).

Shannon, C.E. The Synthesis of Two-Terminal Switching Circuits. *Bell System Tech. J.*, *28*, pp. 59-98. (1949).

Sima, J., and Orponen, P. A computational taxonomy and survey of neural network models. Submitted. (2001).

Siu, K.Y., Roychowdhury, V.P., and Kailath, T. Depth-Size Tradeoffs for Neural Computation *IEEE Transactions on Computers*, *40*, pp. 1402-1412. (1991).

Siu, K.Y., Roychowdhury, V.P., and Kailath, T. *Discrete Neural Computation - A Theoretical Foundation.* Prentice Hall. (1995).

Thornton, C. Parity: the problem that won't go away. In: *Proceedings of AI-96*, Toronto, pp. 362-374. (1996).

Wegener, I. *The complexity of Boolean functions.* Wiley and Sons Inc. (1987).